# Distributed System for Remote Monitoring and Control Greenhouse Environment

**A. Dumitraşcu, D. Ştefănoiu, J. Culiţă, I. Tomiţa**

*Automatic Control and Computers Faculty, Politehnica University of Bucharest Romania (dumalex@ecosys.pub.ro, danny@indinf.pub.ro)*

**Abstract:** This article treats the issue of remote monitoring and control of greenhouse environment through a distributed system that interconnects several subsystems. Thus, a wireless data acquisition subsystem, a fully automated subsystem made of PLCs and industrial communication networks and an irrigation subsystem consists of two water tanks, sensors and actuators are interconnected. According to data provided by sensors, the automatic control subsystem decides whether the irrigation cycle of plants should be performed. In this case, it provides corresponding commands to irrigation subsystem. Moreover, two graphical interfaces (*eKo-View* and *eKo-GreenHouse*) are added of the process and presented in detail in this paper.

## 1. INTRODUCTION

This study of greenhouse microclimate is intended to complete the process of irrigation to keep plants at the appropriate level of development. In our process cycle, the first phase is represented by the acquisition and primary processing of data collected from a wireless sensor network (Ştefănoiu *et al.*, 2010). Central equipment of the sensors network is the eKo-gateway, which collects data and runs a visual monitoring application of the ecological phenomena. Thus, there is a graphical interface that allows the management of parameters and sensors network configuration, called *eKo-View* interface.

eKo-gateway equipment is connected via MPI (MultiPoint Interface) industrial communication bus with PLC S7-300 Simatic class, produced by Siemens®. This is the interconnection between first two subsystems: wireless sensors subsystem and automation control subsystem. The automation control subsystem includes multiple devices (PLCs S7-300 and LOGO!, OP 177B operator board for the local interface with the process parameters). These devices are connected together by different types of dedicated communication bus: Profinet, MPI, AS-I (Actuator and Sensor Interface) and I/O Link. In turn, the automation control subsystem provides commands to the irrigation subsystem to perform proper watering plants. At this level, the actuators such as pomp and electrovalves are controlled (Nitu *et al.*, 2009).

The distributed system can be managed remotely for monitoring and control via a second interface, named *eKo-Greenhouse*. This interface was implemented to create access to the process parameters, actuators, PLC S7-300, export data and log files of events. The two interfaces – *eKo-View* and *eKo-Greenhouse* – offer a series of highly useful capabilities that will be described in section 2. Also, in the next section will be detailed automation solution with all components of the three subsystems. The experimental results of the automated control system are presented in section 3.

## 2. THE AUTOMATION SOLUTION

For the purposes of an efficient implementation the chosen automation solution consists in three subsystems: the primary acquisition and processing of data is performed by a wireless sensors network manufactured by Crossbow® - The *eKo Pro Series* sensors network; the control tasks are performed by a network of Programmable Logic Controllers and interface equipment manufactured by Siemens®. The final set of tasks, the irrigation itself is performed by a third subsystem.

The *eKo Pro Series* wireless sensors network presents itself as a set of nodes (Fig. 1) that wirelessly transmit data collected from sensors attached to them to a radio base. From here, the data gets stored into a single board computer: the eKo-gateway.

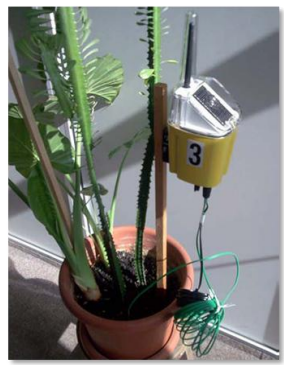

Fig. 1. The wireless node attached to a plant

This data acquisition subsystem is composed of 6 wireless nodes, one for each location with plants in the greenhouse. The eKo-node integrates an IRIS processor/radio board and antenna that are powered by rechargeable batteries and a solar cell. The nodes themselves form a wireless mesh network that can be used to extend the range of coverage. By simply adding an additional eKo-node, it is easy to expand the coverage area. The nodes are pre-programmed and configured with XMesh low-power networking protocol, which provides plug-and-play network scalability for wireless sensor networks.

An eKo-node has four ports to connect the sensors. The system contains a variety of sensors (Fig. 2) such as soil moisture and temperature, ambient temperature and humidity, leaf wetness, soil water content and solar radiation. This solution provides an efficient wireless system for monitoring and acquisition data from multiple locations.

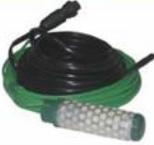

Fig. 2. The sensors of the *eKo Pro* wireless subsystem

In the next figure (Fig. 3) are listed all the ecological parameters that can be monitoring using existing sensors. It is specified for everyone the variation range of measured values, measurement units, and an acronym, which will be used further.

| Soil | Leaf | Greenhouse environment |
|---|---|---|
| *Moisture* (Mo) <br> 0 … 240 [cbar] | *Leaf Wetness* (LeWe) <br> 0 … 1024 [CntS] | *Humidity* (Hu) <br> 0 … 100 [%] |
| *Temperature* (Te) <br> -40 … +65 [°C] | | *Temperature* (Te) <br> -40 … +65 [°C] |
| *Water Contents* (WaCo) <br> 0 … 100 [%wfv] | | *Dew Point* (DwPo) <br> -10 … 50 [°C] |
| | | *Solar Radiation* (SoRa) <br> 0 … 1800 [W/m$^2$] |

Fig. 3. The ecological parameters of the sensors network

The base radio (Fig. 4a) is a fully integrated package that provides the connection between eKo-nodes and sensors and the eKo-gateway. The base radio integrates an IRIS processor/radio board, antenna and USB interface board which is pre-programmed with *XMesh* low-power networking protocol for communication with eKo-nodes. The USB interface is used for data transfer between the base radio and the *eKo-View* interface application running inside the eKo-gateway.

The eKo-gateway (Fig. 4b) runs the *Debian Linux* operating system and comes preloaded with sensor network management and data visualization software packages, *eKo-View* and *XServe*. These programs are automatically started when the gateway is turned on.

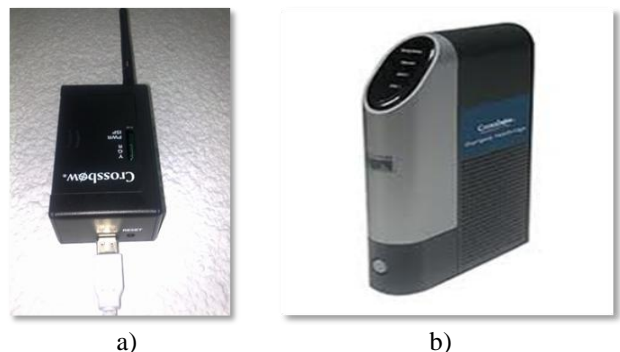

a)                  b)

Fig. 4. The base radio and the eKo-gateway

The data from each sensor is sampled every 15 minutes. The software located in the eKo-gateway then computes average values for every hour, day and month for long term statistics and stores this data into a *SQLite* database that can be exported for processing by third-party applications or simply for backup.

The *eKo-View* application (Fig. 5) offers a familiar and intuitive web browser based interface for sensor network data visualization. In this interface is easy to start monitoring and data acquisition from anywhere in the world via a computer after the sensor network was configured. Through *eKo-View* interface, it is possible to setup and configure the wireless network to display only the data that are interested. The *eKo-View* web interface allows users to make various settings:
- create user-defined map view of sensor nodes across overall network;
- manage user-defined chart configurations;
- create trend charts of multiple sensors across customized time spans;
- view data real-time, which gives users the control needed to manage and maintain crop health;
- view details of individual sensor data;
- monitor performance of network and health of individual nodes;
- set alert levels, run report and get notifications via SMS or email;
- assign custom names to nodes and sensors.

Besides the default functionality presented, additional software was developed to allow the system to communicate the relevant data to the automation equipment: binaries for low level communication via the MPI interface (attached to an USB port of the eKo-gateway) and high level scripts for selecting and sending the data. Therefore, at the level of eKo-gateway equipment have been implemented many applications such as a communication protocol to facilitate data transmission between the eKo-gateway and S7-300 PLC, and PHP scripts that represent the development base of the eKo-Greenhouse interface for remote monitoring and control the process parameters and systems devices.

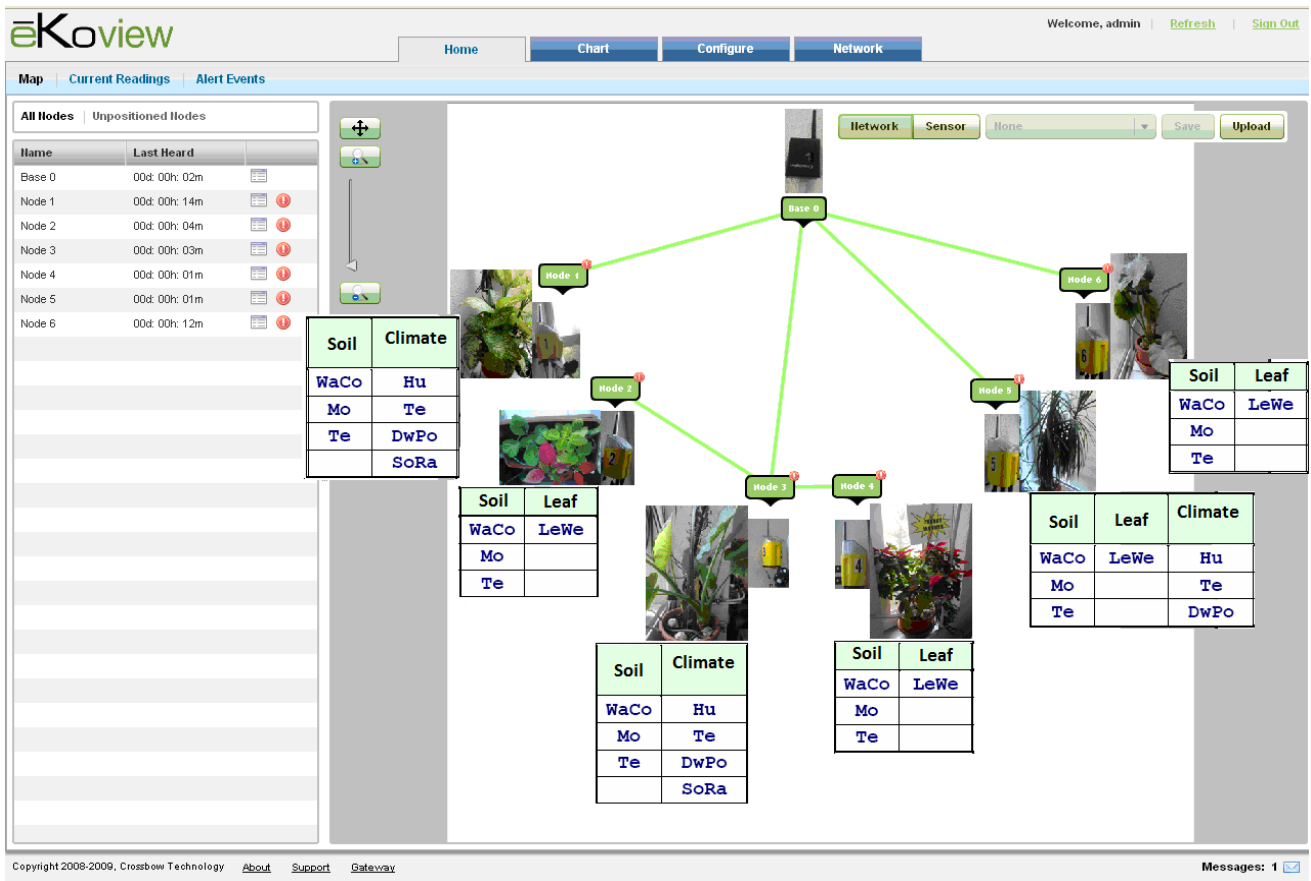

Fig. 5. The wireless sensor network and measured parameters

The control functions of the distributed system reside in the PLC network. An overview of this subsystem can be observed in figure below (Fig. 6).

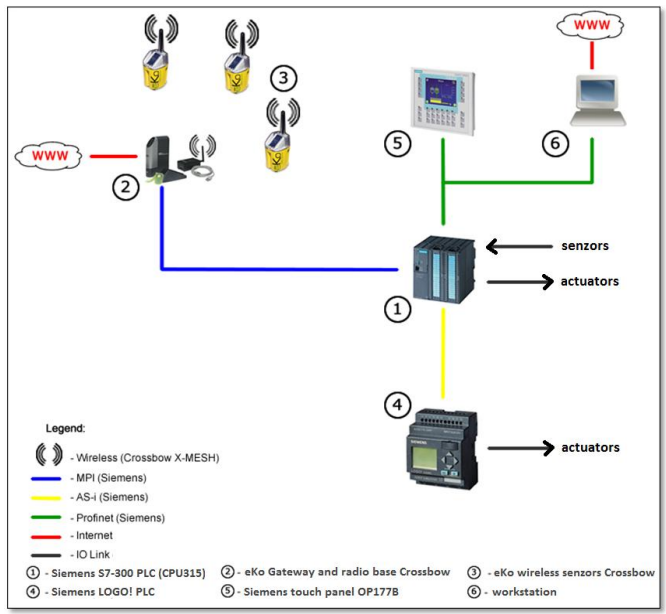

Fig. 6. The control system and its communication

The central unit in the control subsystem is the CPU315F-2DP/PN of the S7-300 Programmable Logic Controller. Here all the data from the *eKo Pro* data acquisition subsystem is collected via the MPI network.

Another industrial communications network used in the control subsystem is the Profinet / Industrial Ethernet network. It is used to accomplish two very important tasks: one is the programming of the CPU315F PLC and the other to transfer data between the CPU and the OP 177B HMI device.

Finally, the interactions with the next subsystem (the irrigations subsystem) are done with the help of a second PLC – a Siemens® LOGO! PLC – that receives its instructions directly from the CPU315F via a AS-Interface network.

The two PLCs interact with the irrigations subsystem with basic I/O digital signals: the water levels in the two irrigation tanks are detected by digital sensors – low, middle and high level for each tank – that are connected directly to the CPU315F. Actuators – the three electrovalves and the pump – are controlled by the LOGO! PLC according to the instructions received from the CPU315F on the AS-interface network.

The most important piece of software in this subsystem is the control software for the CPU315F PLC. This program computes the average values across all the sensors and makes decisions on which of the valves to open so that the humidity values stay within preset limits. Another function of this program is to detect abnormal conditions and defects in the system and take corrective measures. The last of the tasks performed by the central PLC is to log its own actions so that problems are detected and corrected by the human operator.

The next important software resource is the program that runs on the OP177B touch panel. This allows the human operator to have a full view of the system and also to change important parameters in the system like the limits which trigger the opening of valves and the timings for the valves.

To improve system performance, at the level of the control subsystem some measures of reliability are considered. From this point of view, the causes that produce system failure are eliminated.

The irrigation subsystem structure is shown in the following figure (Fig. 7). For the irrigation process two tanks of water are used. The first is a buffer tank (at the bottom of the figure 7), which is fed directly from the mains water supply. The second tank (at the top of the figure 7) is used for the irrigation process of plants.

in an emergency. Also, there is a manual tap with a mash water filter, which is used to retain impurities. Inside the buffer tank are mounted some elements such as a float switch for water supply, three sensors for detecting water level in the tank ("min", "middle" and "max" levels), and a mini-submersible pump. Also, the second tank has three floating level sensors that determine the minimum, middle and maximum levels of water. To fill the second tank with water the mini-pump is used. The irrigation process is performed because the second tank is located at a height of about 3m above the floor.

The irrigation process begins when a normal-closed (NC) type electrovalve receives command to open. This command is sent by the AS-Interface bus from S7-300 PLC to LOGO! PLC after the Master processes the sensors data. Water flows through the pipeline in a time set by user. At the end of the irrigation period, the electrovalve switch in the closed position again (default position).

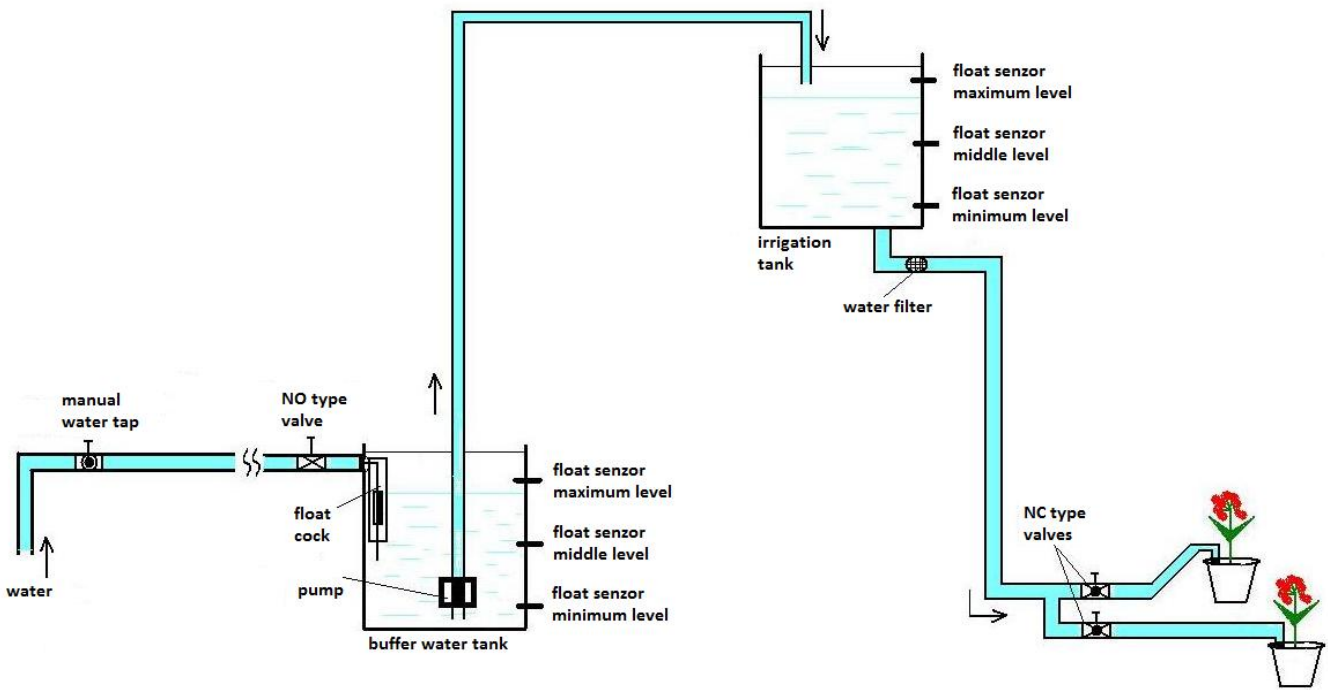

Fig. 7. The structure of the irrigation subsystem.

### 3. THE APPLICATION'S RESULTS

The three subsystems proposed above – data acquisition subsystem, control subsystem, and irrigation subsystem – together form the complete architecture of the process (Dumitraşcu A., 2010), illustrated in the figure 8. This architecture is based on the measurement and data acquisition, management of database, monitoring and data processing, and then the *Master* device (S7-300 PLC) take decisions to control the irrigation system by sending commands to actuators via *Slave* device (LOGO! PLC).

Before the first water tank a normal-open (NO) type electrovalve is installed to interrupt the general water supply

In this architecture a webcam is used 24 hours a day for video monitoring of the process. For this task, another LOGO! PLC is used to control lighting enclosure. Thus, a lamp will be lit at night and goes off next morning.

The web interface to remotely control the automatic irrigation system was built using common web technologies: *HTML* (Hyper Text Markup Language), *JavaScript*, *AJAX* (Asynchronous JavaScript and XML), and *PHP* (PHP Hypertext Processor). This interface is protected. The access is done using a username and a password. Also, the web interface contains the image of process area provided by the webcam. Moreover, in the left side of the interface the last 10 events are precisely shown (date and hour).

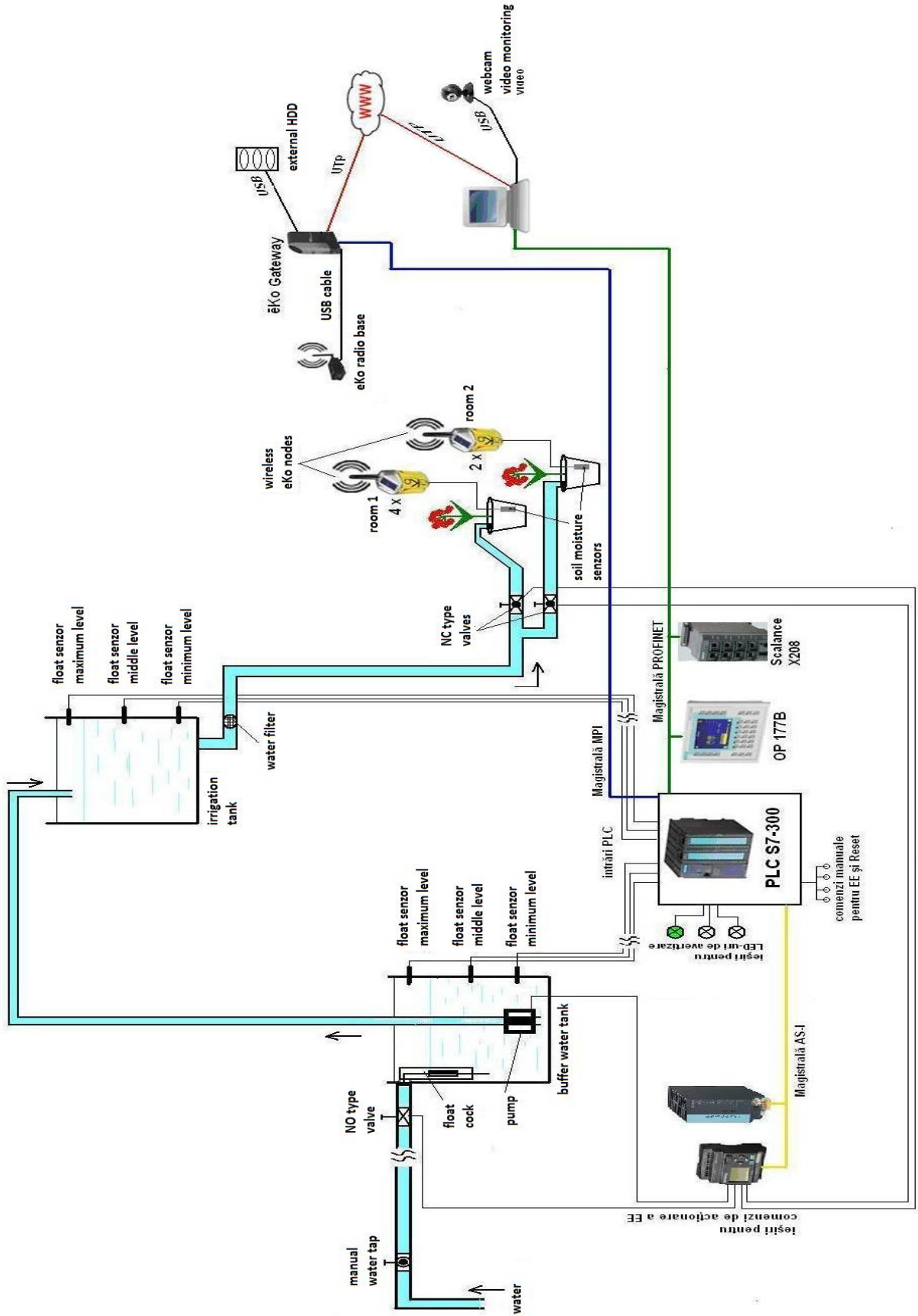

Fig. 8. The structure of the monitoring and control distributed system

The remote control interface of the distributed system is presented in figure 9. The user has the possibility to observe and control the process parameters and system's devices.

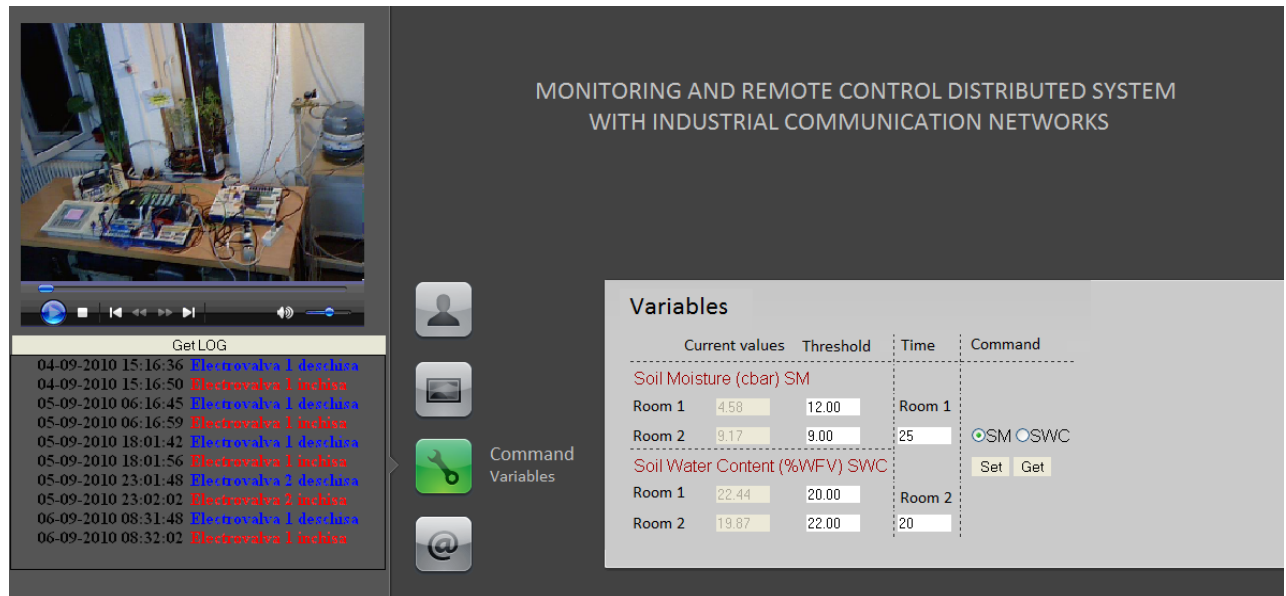

Fig. 9. The web interface for remote process control

In this web interface the control panels are organised into 4 sections, each section containing a set of commands. The first section is shown in figure 10; it contains the commands of S7-300 PLC.

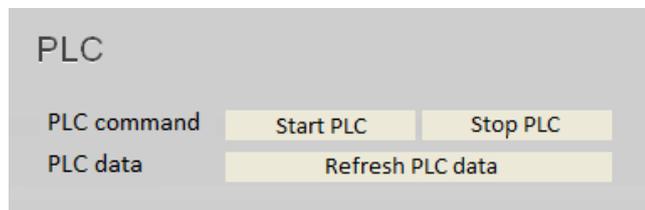

Fig. 10. The S7-300 PLC commands

The second section (Fig. 11) includes remote commands for manual control of actuators (two NC type electrovanes and one pumpe) in irrigation process.

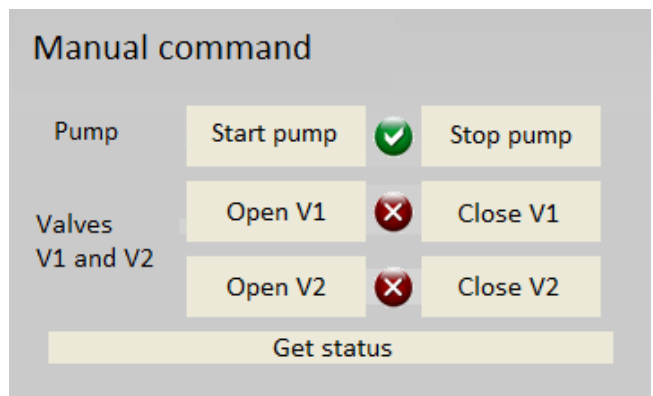

Fig. 11. The commands for manual control of actuators

The next section (Fig. 12) allows user to view and set the application parameters.

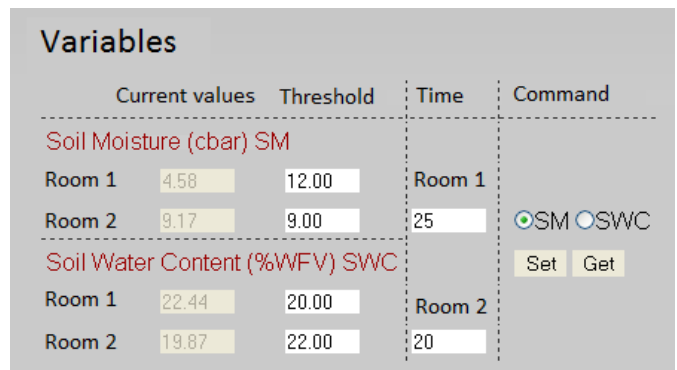

Fig. 12. The interface of application parameters

The last section (Fig. 13) represents an easy modality to export data from the eKo-gateway in *.csv* type file. This opperation is useful for prediction algorithms which consider data as input series.

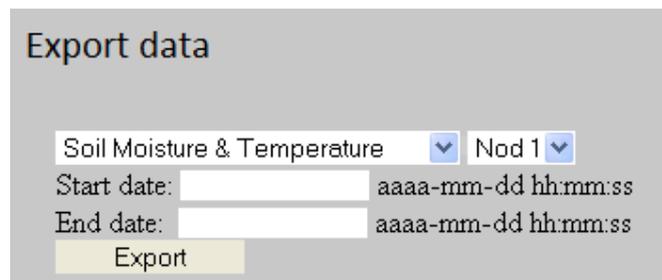

Fig. 13. The section for export data

The system had a great influence over the evolution of the relevant parameter (soil moisture), as best observed in figure 14 that shows the evolution graph for the soil moisture as captured by the sensors planted near the root of the six monitored plants.

The graph is provided by the original software found in the eKo-gateway. The first half represents the period of time in with the irrigation system was turned off. After switching the system on, the variation limits for the soil moisture were severely reduced, translating in better health and plants comfort, which is the purpose of the system, too.

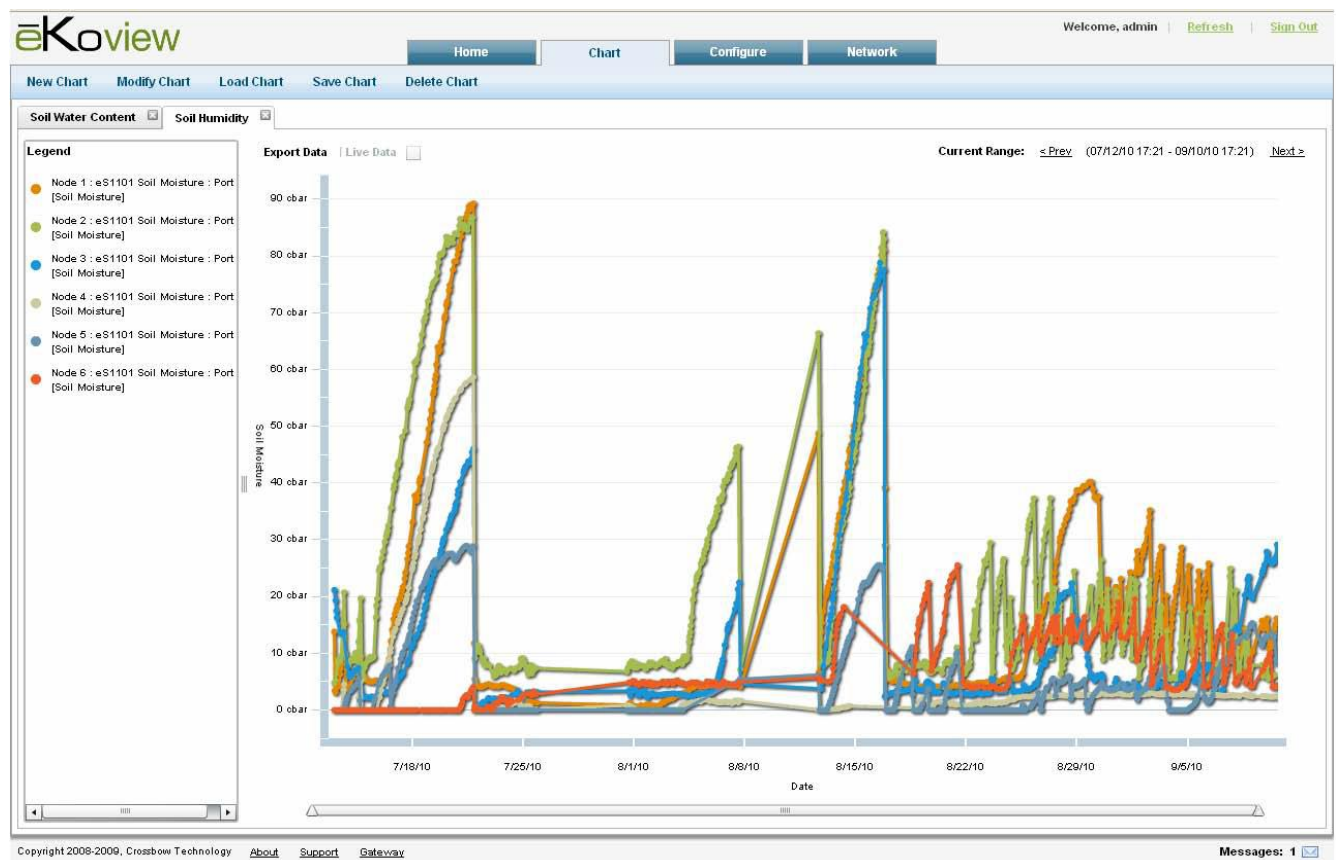

Fig. 14. The experimental results of the irrigation system

## 4. CONCLUSIONS

This paper addresses a very topical subject in the current issue of environment resources, distributed systems, data acquisition and data processing, process control etc. It is therefore submitted a comprehensive application that wants to be an efficient control system of the irrigation process in greenhouse environmental conditions. The implemented application provides a series of operations on process parameters. The first operation is acquisition of data using a wireless sensor network. Another important task is represented by the irrigation process control that is performed in two ways: locally of the process using HMI operator panel and remotely using a web interface that manage equipments and interested parameters' values. All the characteristics of the efficient process control are based on modern equipments, which are current in automation field. Also, regarding the aspect of communication between devices, have been implemented and configured line buses and protocols current used in industry.